# Electromagnetic properties of copper doped lead apatite $Pb_9Cu(PO_4)_6O$

M. Singh[1], P. Saha[1], K. Kumar[2], D. Takhar[3], B. Birajdar[3], V.P.S. Awana[2], S. Patnaik[1*]

[1] *School of Physical Sciences, Jawaharlal Nehru University, New Delhi 110067*

[2] *National Physical Laboratory, K. S. Krishnan Marg, New Delhi*

[3] *Special Centre for Nanoscience, Jawaharlal Nehru University, New Delhi 110067*

Corresponding author spatnaik@jnu.ac.in

**Abstract**

We report on the structural, electrical and magnetic measurements in as-grown polycrystalline samples of $Pb_{10-x}Cu_x(PO_4)_6O$. This compound has been recently reported to be a room temperature superconductor. Our as-grown specimen has excellent XRD matching with the original submission of Lee et al. This sample has 1.5% of $Cu_2S$ as an impurity phase. A resistive transition around 380 K, possibly corresponding to structural transitions of $Cu_2S$, is observed. No evidence of superconducting to normal state transitions in I-V characteristics at room temperature is obtained. Magnetization measurements show linear diamagnetic behaviour that cannot be associated to the superconducting state. Hall measurements provide evidence of hole doping through Cu substitution. In summary, we find no evidence for room temperature ambient pressure superconductivity in Cu doped lead apatite $Pb_9Cu(PO_4)_6O$.

**Introduction**

Over a century since its discovery, the search for a room temperature superconductor continues to be the foremost challenge in material science [1]. In this regard, the establishment of high temperature superconductivity in cuprates, above the liquid nitrogen temperature [2,3], has provided a strong impetus to this movement. But those efforts have saturated by now. Therefore, the recent declaration of possible room temperature superconductivity at ambient pressure in $Pb_{10-x}Cu_x(PO_4)_6O$ [LK-99], at the very least, has rekindled the hope for a room temperature superconductor that can be attained through sustained research [4,5]. Bluntly oblivious to the BCS limit of 23 K, the new Flat band DFT calculations easily reconcile to the idea of a room temperature superconductor [6-8]. Even the age-old broad-band Mott localization ideas, based on resonance valence bond theories, prescribe to such a possibility [1,9]. Of course, the exotic theories proposed to explain the possible superconductivity in $Pb_{10-x}Cu_x(PO_4)_6O$ need a deep scrutiny, but so is the astounding acceptance that the mystery of resistive and magnetic transitions, as reported in LK-99, can be entirely assigned to a less than 5% impurity phase of $Cu_2S$ [10,11]. It is surely worthwhile to ask how a minor impurity phase of $Cu_2S$ can provide percolative current paths and diamagnetic screenings associated to a structural phase transition that can lead to orders of change in electrical resistivity at around 380 K [11]. Furthermore, it will be extremely challenging to proclaim that the observed I-V characteristics in reference 5 and its temperature dependence resembling onset of critical current density, can be associated with physical properties of cuprous sulfide [12]. Last but not the least, to what mechanism can one attribute the splitting of Zero Field and Field cooled DC magnetization protocols (at 10 Gauss field) to the β phase to γ phase structural transition of a robust semiconductor [11]? These are the basic questions that need to be addressed before one can bring in a closure to the mysteries of $Pb_{10-x}Cu_x(PO_4)_6O$.

In specific, one has to be careful to read the experimental data with regard to genuine characterization of the superconducting state. In Wikipedia and elsewhere, a superconducting state is commonly ascribed to a zero electrical resistance state with concurrent onset of diamagnetism. However, experimentally it is impossible to ascertain zero electrical resistance state. The voltage measurements are always limited by resolution of voltmeters. Further, in the nascent stage of processing of new superconductors, the polycrystallinity of samples leads to percolative current paths with negligible inter-grain connectivity. The generally accepted criteria for observation of superconductivity therefore, is the measurement of electric field less than 1μV/cm across the voltage taps. And with the application of magnetic field one would expect the offset of 1 μV/cm criteria to move to lower temperature. Similarly, there is no material which is not diamagnetic; only when other possible magnetic correlations become quiet, diamagnetic susceptibilities come to the fore. The most dependable magnetic signature of superconductivity therefore is splitting of zero field cooled and field cooled magnetization as a function of temperature. The third convincing criteria is the I-V characteristic where on reaching the critical current density, one would expect onset of normal state resistivity.

**Experimental Techniques**

The synthesis of the modified lead apatite $Pb_{10-x}Cu_x(PO_4)_6O$ is mentioned in detail in our earlier work [15]. In brief, first the precursors of $Pb_2SO_5$ (Lanarkite= $PbO + PbSO_4$) and $Cu_3P$ were synthesized using the solid-state reaction. It was followed by the calcination procedure. The precursors were then mixed in $Cu_3P : Pb_2SO_5 = 6:5$ molar ratio to yield $Pb_9Cu(PO_4)_6O$. To ensure homogeneity, the mixture was thoroughly ground and sealed in evacuated quartz tube. The sealed tube was then placed in a furnace and heat treated for 10 hours at 925˚C. After cooling down to room temperature, small chunks of polycrystalline $Pb_9Cu(PO_4)_6O$ were obtained. The detailed structural characterization of the sample was done through powder X-Ray diffraction using Rigaku X-Ray diffractometer. AFM and Raman studies were performed using WItec Alpha 300 RA instrument. High resolution transmission electron microscopy (HRTEM) measurement was done using a JEOL (JEM-2100F) transmission electron microscope. The resistivity measurement has been conducted using the linear four-probe method in the temperature range (215K- 410K). The temperature dependent I-V behaviour have also been analysed for this sample. The nature and concentration of charge carriers were ascertained through Hall measurements.

**Results and Discussions**

Fig. 1(a) shows the X-Ray diffraction pattern of $(Pb_{10-x}Cu_x(PO_4)_6O)$. A few impurity peaks refer to the of synthesized LK99 by-product, namely Pb, $Cu_2S$ and Cu denoted by '+', '*' and '#' respectively. Here, the number of impurity peaks, particularly $Cu_2S$ is lower than the previous reports. The impurity percentage was calculated by summing up the area under impurity peaks, which in turn gives the impurity ratio for $Cu_2S$ ~ 1.5%, Cu ~ 1.6%, and Pb~ 0.2% respectively. Very low impurity ratio gives a clear indication towards sufficient phase purity of obtained LK-99 sample. All the major peaks of this sample match well with the simulated XRD of Apatite structure. The two-phase refinement (considering Apatite as first phase and $Cu_2S$ as second phase) was done to derive the lattice parameters of the LK-99. The obtained lattice parameters are a = b = 9.852(4) Å and c = 7.437(9) Å, which further gives the cell volume to be 721.849 Å$^3$. A minor amount of shrinkage (0.16%) is observed as compared to that of Lead Apatite structure, which denotes successful substitution of Copper atoms to the atomic sites of Lead atoms. Although the shrinkage percent is less than that of reported in [4,5] We note that the major peaks of LK-99 match pretty well with the single crystal data. Schematic diagram of the modified crystal structure (top view) is shown in Figure 1(b). The inner cylindrical structure consists of two opposite facing triangles of Pb1(blue solid spheres) atoms (making hexagonal structure), surrounded by insulating Phosphate ($PO_4$) tetrahedrals. The outer structure of this 3D network contains Pb1 atoms (black solid sphere). When one of the Pb1 atom is replaced by Cu atom (green solid sphere), the stress produced due to structural distortion is transferred to the interface of inner (shown in figure 1(b)) cylinder. This is hypothesized as the source for superconducting quantum well (SQW) generation [4,5]. This condensate is suggested to be confined along c-axis that manifests the 1-D superconductivity. To study the surface morphology of the sample we performed Atomic Force Microscope (AFM) measurement on the sample. AFM micrograph is shown in inset of figure 1(a), which depicts the surface view of the polycrystalline ingot like sample at 1μm Scale. Depth inhomogeneities identified by different colour contrast can be seen in the micrograph. These can be associated to the fact that the sample surface was unpolished. For TEM measurements, powder sample of the $Pb_{10-x}Cu_x(PO_4)_6O$ was prepared by grinding in agate mortar pestle. A suspension of this powder in ethanol solution was prepared by ultrasonication. A drop of this suspension was casted on a holey carbon grid supported by Cu-grid. The TEM bright field image is shown in figure 2(a). The corresponding selected area electron diffraction (SAED) pattern is shown as inset of figure 2(a). First three prominent rings could be indexed as (200), (002), and (210) reflections of the LK-99 phase (with space group $P6_{3/m}$, 176). This is in agreement with the XRD data of LK-99 phase reported recently [4,5]. A corresponding dark field image acquired using a bright spot on the (002) ring (indicated by white dotted ring) is shown in figure 2(b). The dark field image clearly identifies powder particles corresponding to the LK-99 phase.

We also performed the Raman measurements to obtain more structural information about the sample, which is shown in Figure 2(c). The data shows very good resemblance with those reported in [13]. Three Raman spectrums indexed by 1,2 and 3 represents the data taken at three different locations on the sample. It is easy to observe that all the graphs show exactly similar spectra, which depicts the sample to be homogeneous. The main peaks of the spectra correspond to the Phosphate group ($PO_4^{3-}$) and the internal vibrations for these tetrahedral ions are characterized into four types. The peaks shown in Figure 2(c) in the wave number range (340 cm$^{-1}$ – 600cm$^{-1}$ and 870 cm$^{-1}$ to 1080cm$^{-1}$) resembles well with the previously reported vibrations of the Phosphate ($PO_4^{-3}$) group [16]. Initial Peaks (90 cm$^{-1}$ – 230cm$^{-1}$) can be attributed to external vibrations. The main point to be emphasized that some papers [21,22,23] report that the synthesis procedure described in the [4,5] can lead to hydroxide formation. The vibration mode for Hydroxyl ($OH^{-1}$) is attributed to wave number 3565cm$^{-1}$, but our spectra do not show any peak in that wave number range. It rules out the possibility of Hydroxide formation in the as-grown sample.

To test the promise of superconductivity in as-grown LK-99, magnetization measurements were performed. Figure 3(a) shows the MH loop taken at 300K in the field range ±1 Tesla. It shows that at room temperature the sample exhibits diamagnetic but there is no evidence of deviation beyond possible lower critical field ($H_{c1}$). And there is no sign of superconducting hysteresis associated with flux pinning. This is agreement with previous reports [11,15] which implies absence of superconductivity in this copper doped modified Lead apatite. The diamagnetic nature of the sample may be due to the fact that parent lead apatite is diamagnetic. The linear diamagnetic nature in MH curve shows similarity with that magnetization curves of graphite [17]. It is well known that a type of graphite namely Pyrolytic graphite also shows such strong diamagnetism and levitation at room temperature, but it's not superconducting. So, similarities can be drawn for the diamagnetism of LK-99 and that of pyrolytic graphite.

Fig 4(a). depicts the temperature driven change in resistivity in a range of 215K to 410K. The resistivity decreases with increasing temperature. This shows the semiconducting nature of this sample. As the semiconductors have the characteristic energy gap ($E_g$), which is related to the resistivity by the equation

$$\rho = \rho_0 exp^{\frac{E_g}{2k_BT}}$$

By taking the logarithm on both sides it takes the form $ln\rho = ln\rho_0 + E_g/2k_BT$. To determine the semiconducting band gap we plotted the curve between $ln\rho$ vs $T^{-1}$, which is shown in inset of figure 4(a). As the slope of the curve is directly related to energy gap ($E_g$) by the relation $slope = \frac{E_g}{2k_B}$. We get the value of $E_g$ = 0.51eV by substituting the slope's value of 2970.16K. The obtained value of energy gap is less than that of Silicon ($E_g$= *1.1eV*) and Germanium ($E_g$= *0.7eV*). Along with it, no superconducting transition in resistivity is observed in the specified temperature range i.e. (215K- 410K). This gives credence to the theory that the as-grown LK-99 is probably a diamagnetic semiconductor and not a superconductor.

Furthermore, a resistive transition along with a sharp hysteresis is observed between the warming and the cooling curves in the range (360K to 385K). Similar behaviour has been observed in earlier reports [11,14,19] in the same temperature range not only for the LK-99 but also for the $Cu_2S$ pallet. This thermal hysteresis can be safely assigned to a structural transition of $Cu_2S$ [11]. But we note that change in resistance across the structural transition is hardly 2 to 3 times (for 1.5% impurity phase) compared to 4 orders of magnitude change reported by Lee et al.

To calculate the carrier concentration and type of charge carriers, Hall measurement was performed on the sample. Contacts were placed in van der Pauw geometry in 300mT magnetic field. Figure 3(b) shows the variation of Hall voltage with respect to applied current. This gives a slope ($V_H/I$) of value 0.089 which was used to calculate the carrier concentration (*n*) using formula

$$R_H = \frac{1}{ne} = \frac{V_H t}{IB}$$

Here, *t* is the thickness of the sample, that was measured to be 0.15mm. By putting all the values in the formula carrier concentration was estimated to be $1.4 \times 10^{14}$ per cc. The positive slope of the curve depicts that the majority charge carriers in the sample are holes. This hole doping can be understood in the way that in LK-99 with modified lead apatite structure this is achieved by $Cu^{2+}$ substitution at $Pb^{2+}$ site as reported in [4,5]. Pb has $d^{10}$ configuration while Cu has 9 electrons in its d-subshell, hence the Cu substitution makes the LK-99 hole doped. According to Lee et al. [4,5] this copper substitution at a particular Pb site (Pb1) causes the insulator to metal transition (IMT). As Lee et al [5] have discussed, in the superconducting state the resultant potential is sum of two terms namely charge density wave potential ($V_{CDW}$) formed by structural distortion and the critical on-site energy caused by Coloumb interaction ($U_{c, critical\ on\text{-}site\ Coloumb\ energy}$). The second term is affected by two types of structural distortions. One results from the $Cu^{2+}$ substitution in Lead Apatite structure while the second structural change is originated from the difference between the two temperatures, $T_{IMT}$ and $T_c$ respectively. It is suggested that the condensate formation is achieved by the enhanced on-site Coloumb interaction, which leads to superconductivity at room temperature [4,5]. We note that IMT transition can also cause the superconducting transition, but the $T_c$ associated with this transition is not only low but also not of the one-dimensional (depicted in figure 1(c)) characteristic. On the other hand, despite the fact that Cu atoms are effectively substituted, creating the appropriate conditions for IMT and necessary strain to induce the superconductivity, we observe neither IMT nor any superconducting transition in our sample. The as grown LK-99 is semiconducting in the whole temperature range along with high temperature hysteresis observed in the range (360K to 385K). One fact which should be emphasized that according to reference [4,5], the Cu substitution is not the sufficient condition for IMT, Cu atoms should be bonded with the inner Pb cylinder (figure 1(b)) ($3d^9$ subshell of Cu and $6s^2$ subshell of inner Pb atoms) , which is not something that can be controlled just by the reaction parameters and therefore it is not easy to reproduce such selective substitution. A report by Ning Chen et al. by [18] suggests that, in narrow gap semiconductors, thermal fluctuations can affect the carrier concentration in such a way that it can play an important role for hosting the room temperature superconductivity. However, our results significantly differ from the predictions in [18], as the resistivity, gap value and carrier concentrations clearly indicate towards a gapped semiconductor with no signatures of condensate formation.

Inset of figure 4(b) shows the I-V behaviour of this sample at T=300K and 340K. Both the curves exhibit more or less linear characteristic. A very slight deviation from the linear curve can be attributed to the local temperature changes. Since, the sample may contain micro-inhomogeneous phases, which can alter the current paths and local temperature variations. it's not surprising to observe currents switching or non-linearities in I-V curve, which some of the earlier reports have reported [19]. The slope of the curve decreases with increasing temperature. This further illustrates the semiconducting nature of as grown $Pb_{10-x}Cu_x(PO_4)_6O$. This is in direct contrast to the results reported by Lee et al. in reference [5]. There they have provided evidence for sharp onset of normal state resistivity in I-V characteristics as if for the attainment of critical current density. According to report by Kanta et al [20], oxygen occupancy and content can lead to different phase formations which can vary drastically in electronic structure ranging from large gap insulator to metallic states. Along with it these different phase formations can also alter the effect of external impurities like Cu in this case. A possibility of peroxide and superoxide formation is also discussed in [20], that can be negated in our case because peroxide and superoxide formations contain an intrinsic resistance towards hole doping [20]. On the other side, carrier concentration in our sample shows the hole doping.

Furthermore, our attempts to measure the diamagnetic transition across the structural transition range of $Cu_2S$ has not yielded reproducible results. Zhu et al. [11] report a diamagnetic susceptibility of the order of 1E-8 emu/Oe for the ~5% $Cu_2S$ impure sample. It should be pointed out that these are extremely challenging measurements. Deciphering anything less than 1E-6 emu in a VSM attachment of PPMS requires rigorous identification of background and its subsequent subtraction. In contrast, the ZFC -FC split by Lee et al. reports a diamagnetic split of ~ 10E-4 emu/g [5]. This observation of four order of higher diamagnetic susceptibility reported in $Pb_{10-x}Cu_x(PO_4)_6O$ cannot be accounted by 1.5% impurity phase of $Cu_2S$.

**Conclusion**

In summary, we have grown modified lead apatite $Pb_9Cu(PO_4)_6O$ sample with an unintentional ~1.5% of impurity phase of $Cu_2S$. Raman spectra eliminates the possibility of any hydroxide formation. It confirms the phase to be similar with that of reported by Lee et al. The temperature dependent resistivity shows semiconducting nature. A resistive transition is observed at around 380 K that shows pronounced hysteresis under thermal cycling. The current-voltage (I-V) characteristic and magnetic measurement data are in contradiction to the claim of superconductivity in $Pb_9Cu(PO_4)_6O$. From the Hall measurements the carrier concentration in this hole doped semiconductor is estimated to be $1.4\times10^{14}$ per cc. It is therefore safe to conclude that electric transport and magnetic susceptibility data, resembling possible superconductivity in $Pb_9Cu(PO_4)_6O$ in reference 4 and 5, cannot be fully assigned to a minor impurity phase of $Cu_2S$. More studies are needed towards selective site doping in these lanarkite systems that may lead to insulator to metal transition at low dimensions.

## Acknowledgement

M. Singh acknowledges CSIR for providing CSIR-SRF. We acknowledge FIST program of the Department of Science and Technology for the use of CFM system at JNU. We are thankful to AIRF-JNU for providing the facility of Raman spectroscopy, AFM and TEM.

## Data and code availability

The data that supports the results are available upon reasonable request from the corresponding author.

## Author contributions

All authors have contributed to this study. MS: Data curation, analysis, investigation, writing original draft. PS: data curation, investigation, visualization. KK: Sample preparation, data curation. DT: data curation, investigation. BB: writing reviewing and editing. VPSA: Writing-reviewing and editing. SP: Investigation, writing-reviewing and editing, funding acquisition.

## Declarations

**Conflict of interest** The authors declare that there is no conflict of interest.

**Ethical approval** Not applicable.

# References

1. Baskaran, G. (2009). Five-fold way to new high T c superconductors. *Pramana*, *73*, 61-112.

2. Bednorz, J. G., & Müller, K. A. (1986). Possible high T c superconductivity in the Ba−La− Cu− O system. *Zeitschrift für Physik B Condensed Matter*, *64*(2), 189-193.

3. Geballe, T. H. (1993). Paths to higher temperature superconductors. *Science*, *259*(5101), 1550-1551.

4. Lee Sukbae, Ji-Hoon Kim, and Young-Wan Kwon. "The First Room-Temperature Ambient-Pressure Superconductor." *arXiv preprint arXiv:2307.12008* (2023).

5. Lee S., Kim, J., Kim, H. T., Im, S., An, S., & Auh, K. H. (2023). Superconductor $Pb_{10-x}Cu_x(PO_4)_6O$ showing levitation at room temperature and atmospheric pressure and mechanism. *arXiv preprint arXiv:2307.12037*.

6. Cabezas-Escares, J., Barrera, N. F., Cardenas, C., & Munoz, F. (2023). Theoretical insight on the LK-99 material. *arXiv preprint arXiv:2308.01135*.

7. Kurleto, R., Lany, S., Pashov, D., Acharya, S., van Schilfgaarde, M., & Dessau, D. S. (2023). Pb-apatite framework as a generator of novel flat-band CuO based physics, including possible room temperature superconductivity. *arXiv preprint arXiv:2308.00698*.

8. Si, L., & Held, K. (2023). Electronic structure of the putative room-temperature superconductor $Pb_9Cu(PO_4)_6O$. *arXiv preprint arXiv:2308.00676*.

9. Baskaran, G. (2023). Broad band mott localization is all you need for hot superconductivity: Atom mott insulator theory for cu-pb apatite. *arXiv preprint arXiv:2308.01307*.

10. Garisto, D. (2023). LK-99 isn't a superconductor—how science sleuths solved the mystery. *Nature*, *620*(7975), 705-706.

11. Zhu, S., Wu, W., Li, Z., & Luo, J. (2023). First order transition in $Pb_{10-x}Cu_x(PO_4)_6O$ (0.9< x< 1.1) containing $Cu_2S$. *arXiv preprint arXiv:2308.04353*.

12. Hirahara, E. (1951). The physical properties of cuprous sulfides-semiconductors. *Journal of the Physical Society of Japan*, *6*(6), 422-427.

13. Patent Publication 10-2023-0030188 (2023) Title of invention Room-temperature, normal-pressure superconducting ceramic compound and manufacturing method.

14. Wu, H., Yang, L., Yu, J., Zhang, G., Xiao, B., & Chang, H. (2023). Observation of abnormal resistance-temperature behavior along with diamagnetic transition in $Pb_{10-x}Cu_x(PO_4)_6O$-based composite. *arXiv preprint arXiv:2308.05001*.

15. Kumar, K., Karn, N. K., Kumar, Y., & Awana, V. P. S. (2023). Absence of superconductivity in LK-99 at ambient conditions. *arXiv preprint arXiv:2308.03544*.

16. Giera, A., Manecki, M., Bajda, T., Rakovan, J., Kwaśniak-Kominek, M., & Marchlewski, T. (2016). Arsenate substitution in lead hydroxyl apatites: a Raman spectroscopic study. *Spectrochimica Acta Part A: Molecular and Biomolecular Spectroscopy*, *152*, 370-377.

17. Li, Z., Chen, L., Meng, S., Guo, L., Huang, J., Liu, Y., ... & Chen, X. (2015). Field and temperature dependence of intrinsic diamagnetism in graphene: Theory and experiment. *Physical Review B*, *91*(9), 094429.

18. Chen, N., Liu, Y., & Li, Y. (2023). Exploring Room-Temperature Superconductivity in Narrow Energy Gap Semiconductors through Thermally Excited Electrons. *arXiv preprint arXiv:2308.13885*.

19. Liu, C., Cheng, W., Zhang, X., Xu, J., Li, J., Shi, Q., ... & Li, Y. (2023). Phases and magnetism at microscale in compounds containing nominal $Pb_{10-x}Cu_x(PO_4)_6O$. *Physical Review Materials*, *7*(8), 084804.

20. Ogawa, K., Tolborg, K., & Walsh, A. (2023). Models of Oxygen Occupancy in Lead Phosphate Apatite $Pb_{10}(PO_4)_6O$. *ACS Energy Letters*, *8*, 3941-3944.

21. Krivovichev, S. V., & Burns, P. C. (2003). Crystal chemistry of lead oxide phosphates: crystal structures of Pb4O (PO4) 2, Pb8O5 (PO4) 2 and Pb10 (PO4) 6O. *Zeitschrift für Kristallographie-Crystalline Materials*, *218*(5), 357-365.

22. Jiang, Y., Lee, S. B., Herzog-Arbeitman, J., Yu, J., Feng, X., Hu, H., & Bernevig, B. A. (2023). $Pb_9Cu(PO_4)_6(OH)_2$: Phonon bands, Localized Flat Band Magnetism, Models, and Chemical Analysis. *arXiv preprint arXiv:2308.05143*.

23. Krivovichev, S. V. (2023). The crystal structure of $Pb_{10}(PO_4)_6O$ revisited: the evidence of superstructure. *arXiv preprint arXiv:2308.04915*.

Figure Captions:

**Figure 1:** (a) shows the refined XRD pattern of $Pb_9Cu(PO_4)_6O$. Inset (i) of figure1(a) is the AFM micrograph of sample at 1μm scale. (b) shows the schematic top view of modified Lead apatite structure with slight distortion caused by Cu substitution. (c) shows the layout of 1D chain along the c- axis, caused by strong on-site Coloumb interaction.

**Figure 2:** (a) shows the TEM bright field image. Corresponding SAED pattern is shown in inset of figure 2(a). Figure 2(b) shows dark field image acquired using a bright spot on the (002) ring. (c) shows Raman spectra of $Pb_9Cu(PO_4)_6O$ sample taken on three different places on sample surface.

**Figure 3:** (a) shows MH loop taken at 300K in the field range B= ±1 Tesla. (b) depicts the Hall voltage ($V_H$) v/s applied current(I) graph at 300K with B=300mT.

**Figure 4:** (a) shows the resistivity measured in four probe geometry in the temperature range (215K-410K). Inset shows the curve between *lnρ-T⁻¹* (black symbols), which is linearly fitted (red solid line). (b) shows the zoomed view of hysteresis obtained in resistivity warming(blue) and cooling(red) cycle in the temperature range (330K-410K). Inset of figure 4(b) shows the I-V data taken at T = 300 K and T = 340 K.

Fig.1

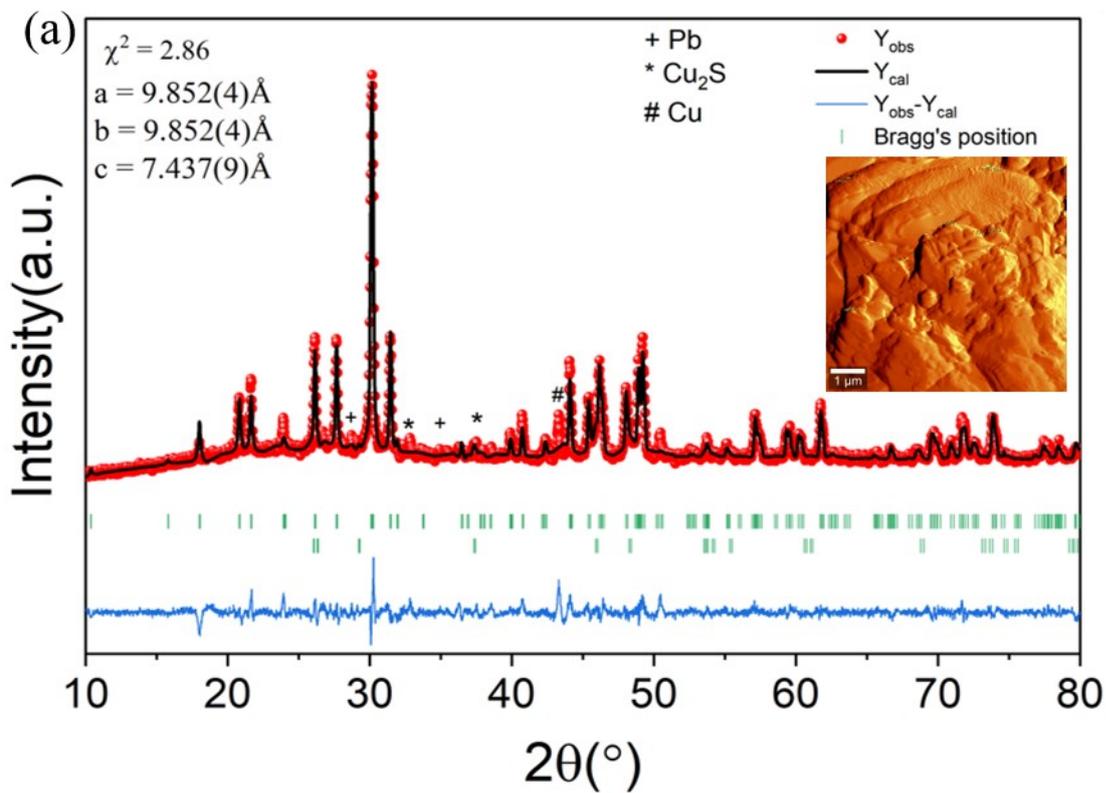

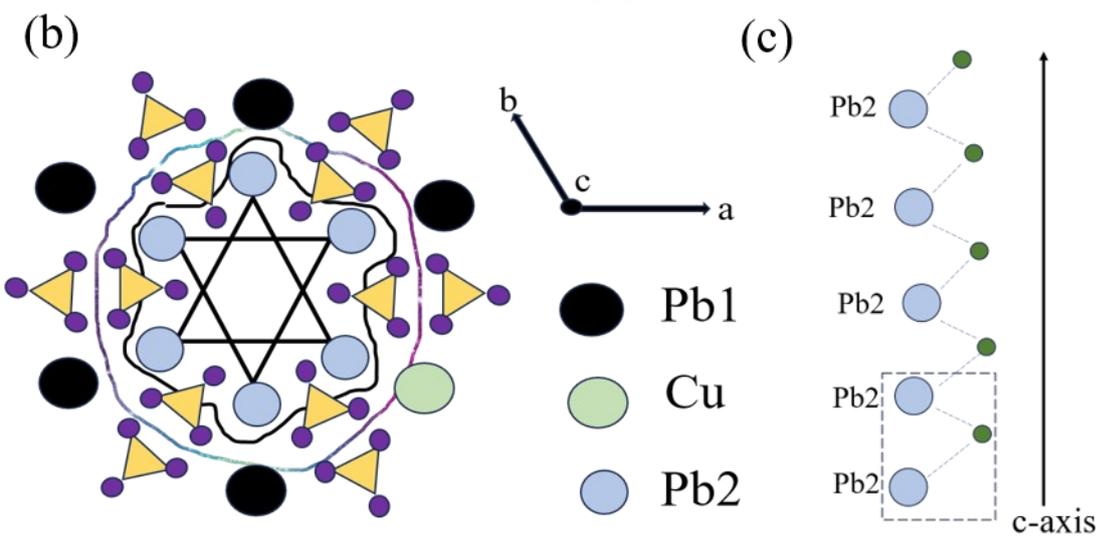

Fig. 2 .

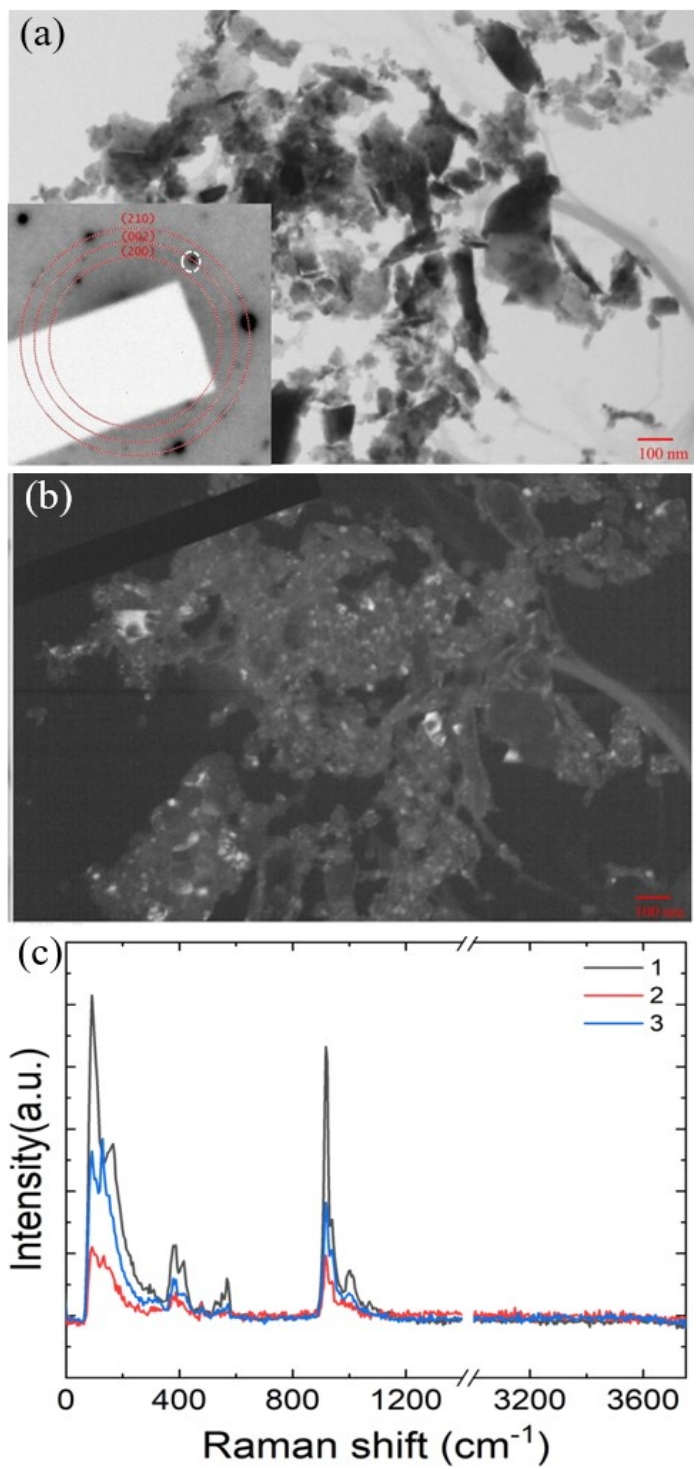

Fig. 3

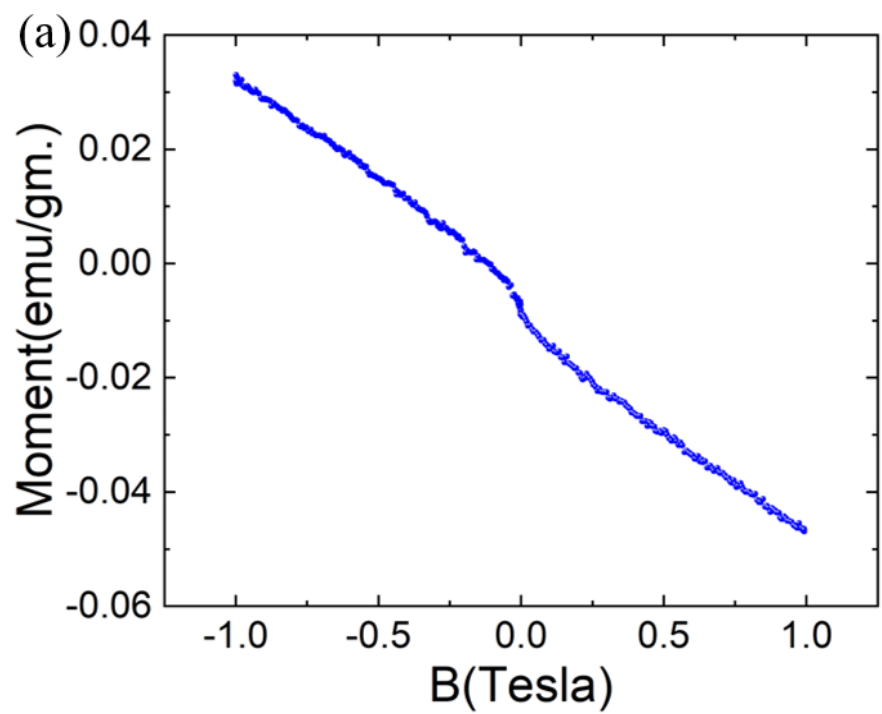

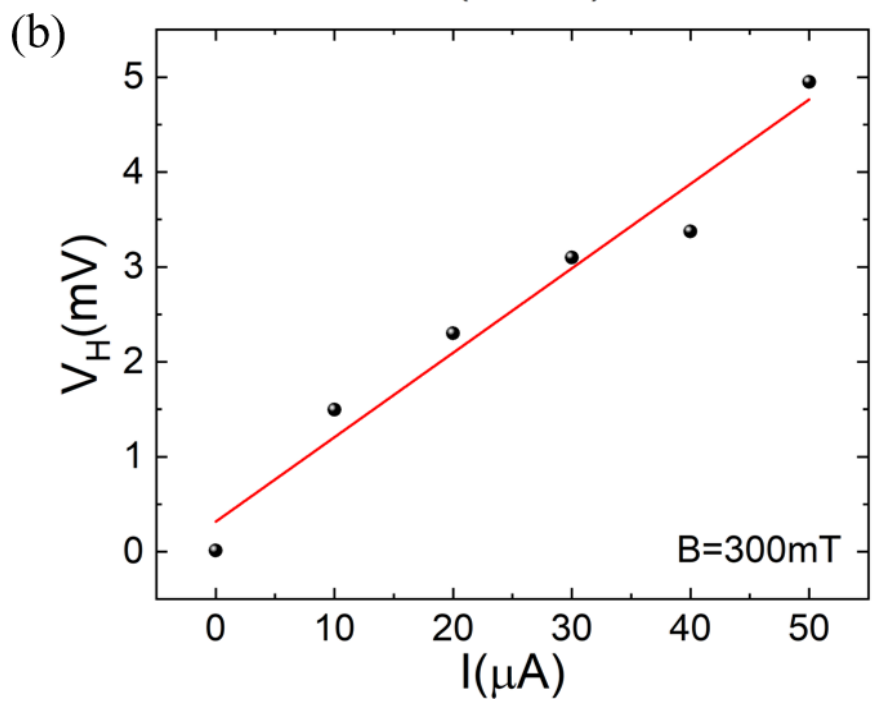

Fig. 4

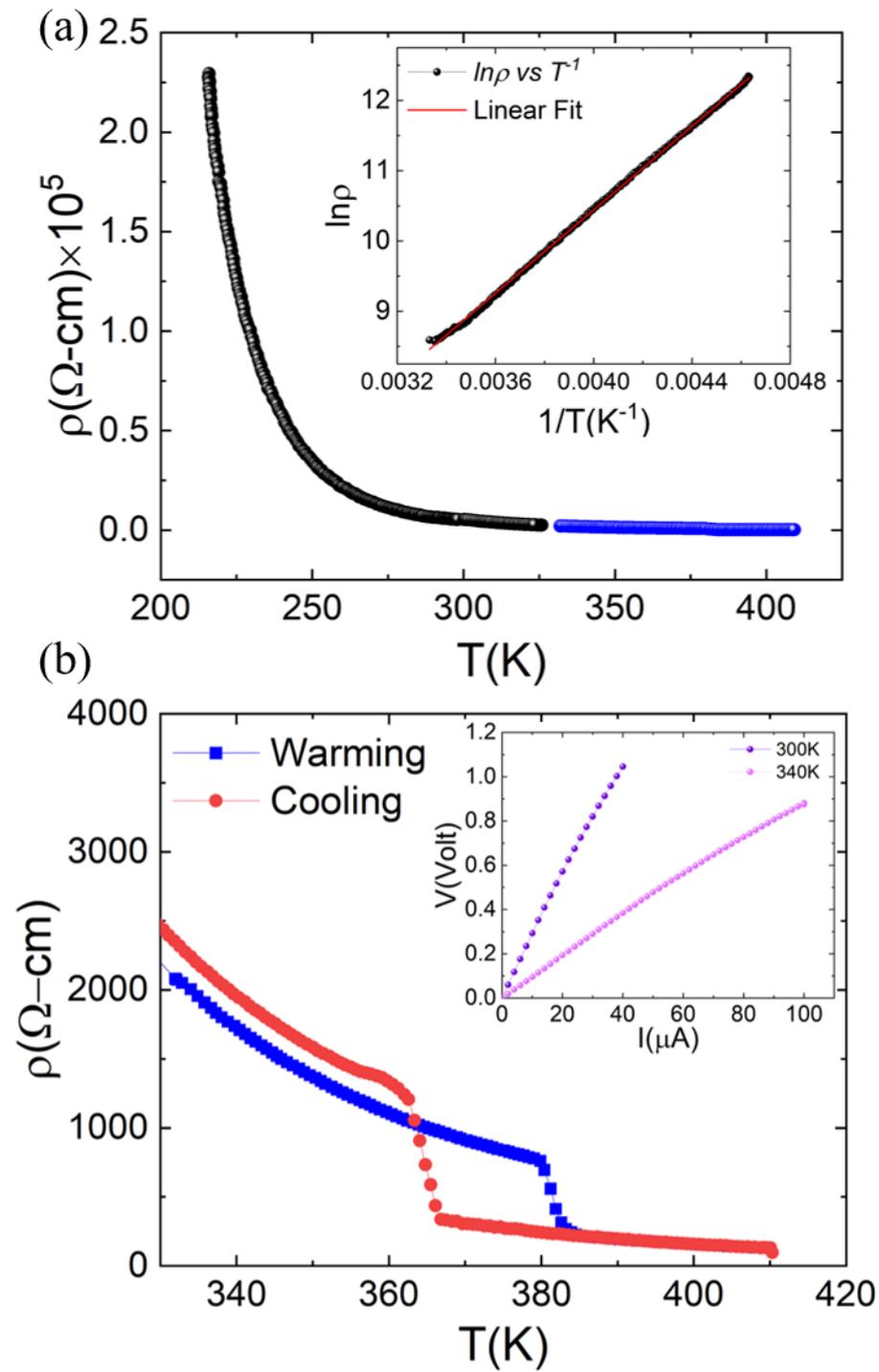